\documentclass[a4paper,12pt]{article}
\usepackage{latexsym}
\usepackage{amssymb}
\usepackage{amsfonts}
\usepackage{amscd}
\newtheorem{defn}{Definition}[section]
\newtheorem{ques}{Question}[section]
\newtheorem{lemm}{Lemma}[section]
\newtheorem{prop}{Proposition}[section]
\newtheorem{theo}{Theorem}[section]
\newtheorem{corl}{Corollary}[section]

\begin{document}

\title{Complete Valuations on \\ Finite Distributive Lattices}
\author{Francesco Marigo}
\date{24 November 2012}
\maketitle

\begin{quote}
ABSTRACT. We characterize the finite distributive lattices which admit a complete valuation, that is bijective over a set of consecutive natural numbers, with the additional conditions of completeness (Definition ˜\ref{Definition:complete}). We prove that such lattices are downset lattices of finite posets of dimension at most two, and determine a realizer through a recursive relation between weights on the poset associated to valuation. The relation shows that the weights count chains in the complementary poset. Conversely, we prove that a valuation defined on a poset of dimension at most two, through the weight function which counts chains in the complementary poset, is complete.
\end{quote}

\section{Introduction}

The starting point of this research was the problem of counting lattice paths. The idea was to think of a lattice path as a line separating a downset of a poset from its complementary upset. In this way the enumeration of lattice paths corresponds to the enumeration of elements of the downset lattice. This can be done assigning a weight to every element of the poset, in such a way that the total weights of downsets are all numbers from $0$ to $n-1$, with $n$ cardinality of the lattice. That is, we define a \emph{valuation} on the lattice, that is bijective over numbers from $0$ to $n-1$. We don't give here an account of the procedure (see, for example of another approach, ˜\cite{fp}). Our purpose is to characterize the finite posets for which a bijective valuation exists. We prove that it exists when the poset has dimension at most two (Proposition ˜\ref{prop:bijec}): it can be defined through a function counting chains in the complementary order. The question (˜\ref{Question:bijec}) if a bijective valuation exists only for posets of dimension two is harder, and we don't prove it. But, by adding a further couple of conditions of regularity, the \emph{completeness} of the valuation (Definition ˜\ref{Definition:complete}), we can prove that such a valuation exists only for posets of dimension at most two (Corollary ˜\ref{Corl:real}). Thus we establish a correspondence between lattices with complete valuations, and posets with couples of linear orders as realizers.

The construction of a complete valuation on a lattice from a realizer of the underlying poset, that we'll give in Section ˜\ref{Section:lexic}, has some points in common with the construction of a pair of diametral linear extensions, as defined, for instance, in ˜\cite{bm} , ˜\cite{fm} , ˜\cite{ma}. The order induced by a complete valuation, which is lexicographic with respect to one of the linear extensions of the poset, is a linear extension of the lattice; the diametrally opposite linear extension would be given by another complete valuation, whose weight function counts chains in the opposite direction. The orders induced by this pair of complete valuation form a revlex pair of linear extensions, as defined in ˜\cite{ma}. There it's proven that a revlex pair is always a diametral pair for downset lattices of posets of dimension two.

\section{Definitions}

Let $(\mathcal{L}, \sqsubseteq , \vee , \wedge )$ be a finite distributive lattice. By \emph{Birkhoff's representation theorem}, $\mathcal{L}$ is isomorphic to the lattice of downsets of a poset $(\mathcal{P},\preceq )$. The correspondence between distributive lattices and posets is one-to-one (up to order isomorphism) so we'll always consider a lattice $\mathcal{L}$ together with the poset $\mathcal{P}$ associated. An element $a \in \mathcal{L}$ corresponds to a downset $A_\downarrow$, generated by a unique antichain $A \subseteq \mathcal{P}$. In order to make the notation lighter, we don't express the correspondence, and we write $a=A_\downarrow$, or $a = x_\downarrow$, with $x \in \mathcal{P}$, when the downset is a lower cone. We'll use the symbols $\cup , \cap$ instead of $\vee , \wedge$ when lattice operations are performed on downsets.

Given $(\mathcal{L},\mathcal{P})$, we'll also consider the dual lattice $\mathcal{L}'$ obtained by reversing the order relation of $\mathcal{L}$. $\mathcal{L}'$ is the lattice of upsets of the same poset $\mathcal{P}$. Similarly, for $b \in \mathcal{L}'$, we write $b=B^\uparrow$, with $B$ antichain. The duality function $\delta : \mathcal{L} \rightarrow \mathcal{L}'$ associates to every downset its complementary upset, so if $b = \delta(a)$, we have $B^\uparrow = \overline{A}_\downarrow$.

We remind ˜\cite{dp} the definition of \emph{valuation} on $\mathcal{L}$, with values in $\mathbb{N}$, the set of natural numbers.

\begin{defn}
A \emph{valuation} is a function $v : L \rightarrow \mathbb{N}$ satisfying:
\begin{itemize}
\item $ v(\bot) = 0 $;
\item $\forall a,b \in \mathcal{L}$, $ a \sqsubseteq b \Rightarrow v(a) \le v(b)$ (monotonicity);
\item $\forall a,b \in \mathcal{L}$, $ v(a \vee b) + v(a \wedge b) = v(a) + v(b) $ (additivity).
\end{itemize}
\end{defn}

It's known ˜\cite{dp} that valuations $v : \mathcal{L} \rightarrow \mathbb{N}$ correspond bijectively to maps $w : \mathcal{P} \rightarrow \mathbb{N}$, setting

\begin{displaymath}
v(a) = \sum_{x \in A_\downarrow} w(x)
\end{displaymath}

We call $w$ the \emph{weight function} associated to the valuation $v$.\\

Given a valuation $v : \mathcal{L} \rightarrow \mathbb{N}$ the dual valuation $v' : \mathcal{L}' \rightarrow \mathbb{N}$ is then defined as

\begin{displaymath}
v'(b) = \sum_{x \in B^\uparrow} w(x)
\end{displaymath}

with the same weight function $w$ associated to $v$.

We define the valuations taking all different values in a set of consecutive numbers.

\begin{defn}
A \emph{bijective valuation} is a valuation $v : \mathcal{L} \rightarrow \mathbb{N}$ with the following additional properties:
\begin{itemize}
\item $v$ takes values in a set $S = \lbrace 0,1,\ldots,n-1 \rbrace$ of consecutive numbers, where $n$ is the cardinality of $L$;
\item $v$ is bijective over $S$.
\end{itemize}
\end{defn}

\begin{ques} \label{Question:bijec}
Which finite distributive lattices admit bijective valuations?
\end{ques}

We aren't able to answer this question in the general case, so we introduce a further condition of regularity, leading to the notion of complete valuation.

Let $\bot$, $\top$ be, respectively, the bottom and top elements of $\mathcal{L}$. We call \emph{initial segment} of $\mathcal{L}$ every subset $\lbrace \bot, a_1,\ldots,a_{i-1} \rbrace \subseteq \mathcal{L}$, with $i \le n$, whose set of $v$ values is the set of numbers $\lbrace 0,1,\ldots,j-1 \rbrace$. We call \emph{final segment} of $\mathcal{L}$ every subset $\lbrace a_j,a_{j+1},\ldots,\top \rbrace \subseteq \mathcal{L}$, with $j \le n$, whose set of $v$ values is the set of numbers $\lbrace j,j+1,\ldots,n-1 \rbrace$.

For every $T \subseteq \mathcal{L}$, we call \emph{join set} of $T$ the set $\bigvee T$ of arbitrary joins of elements of $T$. Similarly, we call \emph{meet set} of $T$ the set $\bigwedge T$ of arbitrary meets of elements of $T$.

\begin{defn} \label{Definition:complete}
A \emph{complete valuation} $v$ is a bijective valuation with the following additional property:
\begin{itemize}
\item for every initial segment $T \subseteq  \mathcal{L}$, the join set $\bigvee T$ is an initial segment (lower completeness);
\item for every final segment $T \subseteq  \mathcal{L}$, the meet set $\bigwedge T$ is a final segment (upper completeness).
\end{itemize}
\end{defn}

As last definition, we remind the notion of \emph{dimension} of a poset ˜\cite{tr} .

\begin{defn}
A poset $\mathcal{P}$ has \emph{dimension at most $d$}, $dim(\mathcal{P})\le d$, if there is a family $R$ of $d$ linear extensions of $\mathcal{P}$
\begin{displaymath}
R=\lbrace \Lambda_1 ,\Lambda_2 ,\ldots,\Lambda_d \rbrace
\end{displaymath}
such that:
\begin{itemize}
\item for every $x,y \in \mathcal{P}$, if $x\preceq y$, then $x\preceq_i y$ in every $\Lambda_i$;
\item for every $x,y \in \mathcal{P}$, if $x$ and $y$ are incomparable, then $x\preceq_i y$ in at least one $\Lambda_i$, and $y\preceq_j x$ in at least one $\Lambda_j$, with $i \ne j$.
\end{itemize}
\end{defn}

$R$ is called a \emph{realizer} of $\mathcal{P}$. If $d$ is the least number such that  $dim(\mathcal{P})\le d$, then  $dim(\mathcal{P}) = d$.

If  $dim(\mathcal{P})\le 2$, then $\mathcal{P}$ has a realizer $R=\lbrace \Lambda_1 , \Lambda_2 \rbrace$ ($\Lambda_1 = \Lambda_2$ when P is a linear order) and there are other three posets associated, by reversing one or two of the linear orders of the realizer. The four cases are:

\begin{itemize}
\item $\mathcal{P}$ with realizer $R=\lbrace \Lambda_1 , \Lambda_2 \rbrace$;
\item $\mathcal{P}'$ with realizer $R'=\lbrace \Lambda'_1 , \Lambda'_2 \rbrace$;
\item $\mathcal{Q}$ with realizer $S=\lbrace \Lambda_1 , \Lambda'_2 \rbrace$;
\item $\mathcal{Q}'$ with realizer $S'=\lbrace \Lambda'_1 , \Lambda_2 \rbrace$.
\end{itemize}

We say that the couple $(\mathcal{Q},\mathcal{Q}')$ is \emph{complementary} to the couple $(\mathcal{P},\mathcal{P}')$.

\section{From valuations to realizers}

Given a lattice $\mathcal{L}$ with a complete valuation $v$, we define on the poset $\mathcal{P}$ two linear orders. Let $\Lambda_\downarrow, \Lambda^\uparrow$ be the following orders:

\begin{itemize}
\item $\Lambda_\downarrow$: $\forall x,y \in \mathcal{P}$, $x \preceq_\downarrow y \Leftrightarrow v(x_\downarrow) \le v(y_\downarrow)$;
\item $\Lambda^\uparrow$: $\forall x,y \in \mathcal{P}$, $x \preceq^\uparrow y \Leftrightarrow v'(y^\uparrow) \le v'(x^\uparrow)$.
\end{itemize}

The orders $\Lambda_\downarrow, \Lambda^\uparrow$ are well defined and they are linear, because the values of each of $v$ and $v'$ are all different.

\begin{theo} \label{Theorem:real}
The set $R = \lbrace \Lambda_\downarrow, \Lambda^\uparrow \rbrace$ is a realizer of $\mathcal{P}$.
\end{theo}

By definition of dimension, we have the following corollary.

\begin{corl} \label{Corl:real}
For any finite distributive lattice $\mathcal{L}$ and complete valuation $v$ on $\mathcal{L}$, the corresponding poset $\mathcal{P}$ through Birkhoff duality has dimension
\begin{displaymath}
dim(\mathcal{P}) \le 2.
\end{displaymath}
\end{corl}

To prove Theorem ˜\ref{Theorem:real} we need some preliminary lemmas.

\begin{lemm}
For every $x,y \in \mathcal{P}$, the following equalities hold:
\begin{equation} \label{Equation:union}
v(y_\downarrow) = 1+ v \left( \bigcup_{x\prec_\downarrow y} x_\downarrow \right);
\end{equation}
\begin{equation} \label{Equation:unionbis}
v'(x^\uparrow) = 1+ v' \left( \bigcup_{x\prec^\uparrow y} y^\uparrow \right).
\end{equation}
\end{lemm}

\emph{Proof.} We prove (˜\ref{Equation:union}) , then (˜\ref{Equation:unionbis}) follows by duality. Let's consider the set

\begin{displaymath}
V = \lbrace 0, 1, 2, \ldots , v(y_\downarrow) \rbrace
\end{displaymath}

and the corresponding, with respect to $V$, sequence of lattice elements

\begin{displaymath}
T = \lbrace \bot , A_\downarrow^1 , A_\downarrow^2 , \ldots , y_\downarrow \rbrace.
\end{displaymath}

All the $x_\downarrow$ such that $x\prec_\downarrow y$ belong to $T$, and for each element $A_\downarrow \in T$, if $x\in A_\downarrow$, it must hold $x \prec_\downarrow y$. Therefore

\begin{displaymath}
\bigcup_{x\prec_\downarrow y}x_\downarrow = \bigcup_{A_\downarrow \in T}A_\downarrow .
\end{displaymath}

For lower completeness, the join set of $\bigvee T$ is an initial segment, and let $k$ be its highest value. Since no join of elements of $T$ can be equal to $y_\downarrow$ (because $y$ does not belong to any element of $T$), it must be $k<v(y_\downarrow)$, hence

\begin{displaymath}
1+ v \left( \bigcup_{x\prec_\downarrow y} x_\downarrow \right) \le v(y_\downarrow).
\end{displaymath}

If $z \in A_\downarrow$, with $y\preceq_\downarrow z$, then $v(y_\downarrow)\le v(A_\downarrow)$, so $v(y_\downarrow)$ is the least value of $v$ on downsets containing elements $z$ with $y\preceq_\downarrow z$. It follows that the previous inequality is actually an equality.
\hfill $\Box$

\begin{lemm} \label{Lemma:sums}
For every $x,y \in \mathcal{P}$, it holds:
\begin{equation} \label{Equation:sum}
w(y) = 1+ \sum_{x\prec_\downarrow y , x\nprec y} w(x);
\end{equation}
\begin{equation} \label{Equation:sumbis}
w(x) = 1+ \sum_{x\prec^\uparrow y , x\nprec y} w(y).
\end{equation}
\end{lemm}
\emph{Proof.} For transitivity of the order relation, we have
\begin{displaymath}
\bigcup_{x\prec_\downarrow y}x_\downarrow = \bigcup_{x\prec_\downarrow y} \lbrace x \rbrace.
\end{displaymath}
Hence
\begin{displaymath}
v \left( \bigcup_{x\prec_\downarrow y}x_\downarrow \right) = \sum_{x\prec_\downarrow y} w(x) = \sum_{x\prec_\downarrow y ,  x\nprec y} w(x) + \sum_{x\prec y} w(x).
\end{displaymath}
On the other hand,
\begin{displaymath}
v (y_\downarrow) = \sum_{x\preceq y} w(x) = w(y) + \sum_{x\prec y} w(x).
\end{displaymath}
Then (˜\ref{Equation:sum}) follows from last two equalities and (˜\ref{Equation:union}). (˜\ref{Equation:sumbis})  follows by duality from (˜\ref{Equation:unionbis}).
\hfill $\Box$

\emph{Proof (of Theorem ˜\ref{Theorem:real}).} We have to prove the following implication:
\begin{equation} \label{Equation:order}
\forall x,y \in \mathcal{P}, \lbrack x\prec y \Leftrightarrow (x\prec_\downarrow y , x\prec^\uparrow y)\rbrack .
\end{equation}

If $x\preceq y$, then $x_\downarrow \subseteq y_\downarrow$ and $y^\uparrow \subseteq x^\uparrow$, hence $v(x_\downarrow) \le v(y_\downarrow)$ and $v'(y^\uparrow) \le v'(x^\uparrow)$, which proves the right implication by definition of $\Lambda_\downarrow$ and $\Lambda^\uparrow$.

The left implication is proved by the following lemma.
\begin{lemm} \label{Lemma:contr}
The following implications hold, for every $x,y \in \mathcal{P}$:
\begin{equation}
\lbrack x\prec_\downarrow y, x \nprec y \rbrack \Rightarrow w(x)<w(y);
\end{equation}
\begin{equation}
\lbrack x\prec^\uparrow y, x \nprec y \rbrack \Rightarrow w(y)<w(x).
\end{equation}
\end{lemm}
\textbf{Proof.} It follows from (˜\ref{Equation:sum}) and (˜\ref{Equation:sumbis}).
\hfill $\Box$

The proof of Theorem ˜\ref{Theorem:real} is then complete, because Lemma ˜\ref{Lemma:contr} implies that
\begin{displaymath}
( x\prec_\downarrow y, x\prec^\uparrow y ) , x \nprec y
\end{displaymath}
is a contraddiction.
\hfill $\Box$

The recursive relations of Lemma ˜\ref{Lemma:sums} give us an interpretation of weights as enumeration of antichains, or chains of the complementary order.

Let $\preceq '$ be the complementary order relation obtained by reversing $\Lambda^\uparrow$,
\begin{displaymath}
x\preceq ' y \Leftrightarrow (x\preceq_\downarrow y , y\preceq^\uparrow x).
\end{displaymath}

\begin{prop}
For every $y \in \mathcal{P}$, it holds:
\begin{equation} \label{Equation:sumrev}
w(y) = 1+ \sum_{x\prec ' y} w(x).
\end{equation}
\end{prop}

\emph{Proof.} From (˜\ref{Equation:order}) we have

\begin{displaymath}
\lbrack x \prec_\downarrow y , x\nprec y \rbrack \Leftrightarrow ( x \prec_\downarrow y , y \prec^\uparrow x ),
\end{displaymath}

therefore the equation (˜\ref{Equation:sumrev}), with respect to the weight func equals the sum in (˜\ref{Equation:sum}).
\hfill $\Box$

\begin{corl}
$w(y)$ is the number of chains in $\mathcal{Q}$ (the complementary poset, with order relation $\preceq'$ ) with $y$ as maximum element.
\end{corl}

\emph{Proof.} It's easy to see that the number of chains in $\mathcal{Q}$ with maximum $y$ satisfies the same relation as (˜\ref{Equation:sumrev}). In fact, for every maximum $y$, there are one chain with the only $y$, and one chain for every chain with maximum $x$ such that $x\prec ' y$, obtained by adding $y$ as maximum.
\hfill $\Box$

In the next section we'll reverse the problem, starting from a weight function which counts chains in the complementary order, and proving that the corresponding valuation is complete.

\section{From realizers to valuations} \label{Section:lexic}

Given a poset $\mathcal{P}$ of dimension $dim(\mathcal{P})\le 2$, with a realizer $\lbrace\Lambda_1 , \Lambda_2 \rbrace$, we define a weight function $w:\mathcal{P}\rightarrow\mathbb{N}$ as the function counting chains in the complementary poset $\mathcal{Q}$ with realizer $\lbrace\Lambda_1 , \Lambda'_2 \rbrace$:

\begin{displaymath}
w(x) = \vert \lbrace \gamma = (x_1 ,\ldots , x_k , x) : x_1\prec ' \ldots \prec ' x_k \prec ' x\rbrace \vert.
\end{displaymath}

\begin{prop} \label {prop:bijec}
The valuation $v:\mathcal{L}\rightarrow\mathbb{N}$ associated to $w$ is bijective.
\end{prop}

\emph{Proof.} It's easy to see that $v(\top)=n-1$. In fact the sum of all values of $w$ is the total number of chains in $\mathcal{Q}$, that is the number of antichains in $\mathcal{P}$, minus one, the empty chain.

Since $v$ attains the values $0$ and $n-1$, it's sufficient to prove that $v$ is surjective. We do it by induction on $i$ from $0$ to $n-1$. Let $A_i$ be such that $v(A_{i \downarrow })=i$, and $B_i$ such that $B_i^\uparrow = \overline{A}_{i \downarrow }$. $B_i$ is a chain in $\mathcal{Q}$, therefore it has a minimum element $y=min_\mathcal{Q} B_i$. The set

\begin{displaymath}
Z=\lbrace x | x \preceq_1 y \rbrace
\end{displaymath}

is a downset, hence

\begin{displaymath}
Y = (  A_{i \downarrow } \cup y_\downarrow ) \cap Z
\end{displaymath}

is a downset too. $Y$ can be written

\begin{displaymath}
Y = ( A_{i \downarrow } \cup  \lbrace y \rbrace ) \setminus \lbrace x | x \prec ' y \rbrace = A_{i+1 \downarrow },
\end{displaymath}

and, by the recursive relation ˜\ref{Equation:sumrev}, it holds

\begin{displaymath}
v (A_{i+1 \downarrow }) = v (A_{i \downarrow }) + w(y) - \sum_{x\prec ' y} w(x) = i+1 .
\end{displaymath}

Then surjectivity follows by induction.
\hfill $\Box$

To prove the completeness of $v$, we need a preliminary construction. Let $x_j$ be the element of position $j$ in the order $\Lambda_1$. We define a function $\Omega : \mathcal{L}\rightarrow\lbrace 0,1\rbrace^k$:

\begin{displaymath}
\Omega (A_\downarrow) = (d_1 , d_2 , \ldots , d_k)
\end{displaymath}

where $k = \vert \mathcal{P}\vert$ and $d_j \in \lbrace 0,1 \rbrace$, such that

\begin{displaymath}
d_j = 1 \Leftrightarrow x_j \in A_\downarrow .
\end{displaymath}

We consider now the elements $a_i = A_{i \downarrow }$ ordered with respect to $v$, as in the proof of Proposition ˜\ref{prop:bijec}.

\begin{prop} \label{prop:lexi}
The elements $a_i \in \mathcal{L}$ are ordered lexicographically with respect to $\Omega$.
\end{prop}

\emph{Proof.} At each inductive step of the proof of Proposition ˜\ref{prop:bijec}, the element added, $y$, is greater, in the order $\Lambda_1$, of any element of the set removed, $\lbrace x | x \prec ' y \rbrace$. Therefore $a_{i+1}$ is greater than $a_i$, in the lexicographic order induced by $\Omega$.
\hfill $\Box$

We can now state a theorem, which plays the role of inverse of Theorem˜\ref{Theorem:real}.

\begin{theo} \label{Theorem:comp}
The valuation $v:\mathcal{L}\rightarrow\mathbb{N}$ associated to $w$ is complete.
\end{theo}

\emph{Proof.} Let

\begin{displaymath}
T= \lbrace \bot , a_1 , \ldots , a_h \rbrace
\end{displaymath}

be an initial segment. let $m$ be the highest value such $x_m \in T$ (as before, the index stands for the order $\Lambda_1$ in $\mathcal{P}$). Since $T$ is lexicographically ordered, no $a_i$ in $T$ contains $x_{m+1}$. Therefore the join set $\bigvee T$ equals the join set of

\begin{displaymath}
T'= \lbrace \bot , x_1 , \ldots , x_j \rbrace.
\end{displaymath}

$\bigvee T'$ is the set of elements of $\mathcal{L}$ containing only elements $x_j \in \mathcal{P}$ with $j \le m$. These are all the elements of $\mathcal{L}$ preceding $x_{m+1 \downarrow }$ lexicographically, therefore, for Proposition ˜\ref{prop:lexi} they precede $x_{m+1 \downarrow }$ also in the order induced by the valuation, that is they form an initial segment. This proves lower completeness; for upper completeness the proof follows the same lines by duality.
\hfill $\Box$

Theorem ˜\ref{Theorem:real} and Theorem ˜\ref{Theorem:comp} can be summarized in the following, that is the final result of this paper.

\begin{theo}
There is a bijection between finite distributive lattices $\mathcal{L}$ with complete valuations $v: \mathcal{L} \rightarrow \mathbb{N}$, and posets $\mathcal{P}$ with realizers $R=\lbrace\Lambda_1 ,\Lambda_2 \rbrace$,
\begin{displaymath}
(\mathcal{L} , v) \longleftrightarrow (\mathcal{P} ,\Lambda_1 ,\Lambda_2).
\end{displaymath}
The bijection is Birkhoff duality if restricted to $\mathcal{L}$ and $\mathcal{P}$. The weight function $w: \mathcal{P} \rightarrow \mathbb{N}$ associated to $v$ has value:
\begin{displaymath}
w(x) = \vert \lbrace \gamma = (x_1 ,\ldots , x_k , x) | x_1\prec ' \ldots \prec ' x_k \prec ' x\rbrace \vert ,
\end{displaymath}
where $\prec '$ is the order relation of the poset $\mathcal{Q}$ of realizer $S=\lbrace \Lambda_1 ,\Lambda'_2 \rbrace$.
\end{theo}

\end{document}